\documentclass[twocolumn,prl,floatfix]{revtex4}

\usepackage[dvips]{graphicx}
\newcommand{\beq}{\begin{equation}}
\newcommand{\eqref}[1]{Eq.\ (\ref{#1})}
\newcommand{\refref}[1]{Ref.\ \onlinecite{#1}}
\newcommand{\enq}{\end{equation}}   
\newcommand{\dirl}{{i}}   % parallel index
\newcommand{\dirn}{{\ell}}         % layer (transverse) index
\newcommand{\dirm}{{j}}   % parallel nbr
\newcommand{\dirp}{{m}}            % transverse nbr index

\newcommand{\la}{\langle}
\newcommand{\ra}{\rangle}

\begin{document}

\title{
Driven depinning of strongly disordered media and
anisotropic mean-field limits}
\date{\today}

\author{M. Cristina Marchetti, A. Alan Middleton, Karl Saunders, and
  J. M. Schwarz}
\affiliation{Physics Department, Syracuse University, Syracuse, NY 13244}

\begin{abstract} 
Extended systems driven through strong disorder are modeled 
generically using coarse-grained degrees of freedom that interact
elastically in the directions parallel to the driving force and that
slip along at least one of the directions transverse to the motion. A
realization of such a model is a collection of elastic channels with
transverse viscous couplings. In the infinite range limit this model
has a tricritical point separating a region where the depinning is
continuous, in the universality class of elastic depinning, from a
region where depinning is hysteretic. Many of the collective transport
models discussed in the literature are special cases of the generic
model.
\end{abstract}  


\maketitle  

%\section{Introduction} 

Nonequilibrium transitions from stuck to moving states underlie the
physics of a wide range of phenomena, from fracture and earthquake
rupture to flux flow in type-II superconductors \cite{Fisher98}.
Various dynamical models have been proposed in different contexts.
One class of models, overdamped {\it elastic} media pulled by an
applied force $F$, has been studied extensively.  These predict a
nonequilibrium phase transition from a pinned state to a sliding state
at a critical value $F_T$ of the driving force. Generically, this
depinning transition displays critical behavior as in equilibrium {\it
continuous} transitions \cite{footcontinuous}, with the medium's mean
velocity $v$ acting as the order parameter.  In overdamped elastic
media, the sliding state is unique and no hysteresis can
occur \cite{AAMunique}. Universality classes have been identified,
which are distinguished, for example, by the range of interactions or
by the periodicity (or nonperiodicity) of the pinning forces.
%Condensed matter systems that
%may exhibit this behavior range from CDWs in anisotropic metals to
%crack fronts in brittle materials.

The elastic medium model is often inadequate to describe many systems
which exhibit plasticity (e.g., due to topological defects in the
medium) or inertial effects (underdamping.)  The dynamics in plastic
systems can be inhomogeneous, with coexisting pinned and moving
regions.  The depinning transition may be discontinuous (first order),
possibly with macroscopic hysteresis.  Several mean-field models of
driven extended systems with locally underdamped relaxation or phase
slip have been proposed in the literature
\cite{strogatz88&89,levy92,levy94,NV97,Fisher98,MMP00,SF01}.

In this paper, we present a model of driven disordered systems that
encompasses several of the models discussed in the literature.  This
model incorporates the anisotropy of the sliding state in the plastic
flow region that results either from flow along coupled channels
oriented in the direction of drive (e.g., as in the moving smectic
phase \cite{BMCMR97}) or in layered materials such as the high-$T_c$
cuprate superconductors.  We restrict ourselves to systems with a
periodic structure along the direction of motion, such as charge
density waves (CDWs), 2D colloidal arrays, and vortex lattices in type-II
superconductors and consider only the dynamics of a scalar
displacement field.  This model uses coarse-grained degrees of freedom
that are solid-like regions.  These volumes slip relative to each
other in the transverse dimensions, due to the presence of small scale
defects (phase slips, dislocations, grain boundaries) at their
boundaries, but remain elastically coupled in the longitudinal
dimensions.  Our current results are for the mean-field limits, in the
transverse dimensions at least, with transverse viscous interactions.
The studies carried out so far of these types of models for finite
range interactions \cite{MCMKD02,SF02} suggest that the mean field
approximation described here may give the correct topology for the
phase diagram, although there will be corrections to the critical
behavior in finite dimensions.

\begin{figure}
\centering
\includegraphics[width=8.cm]{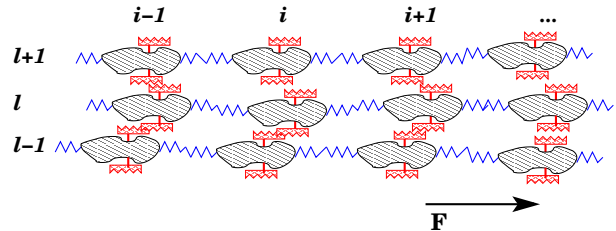}
\caption{\label{fig:cartoon} A two-dimensional realization of the
anisotropic driven medium described in the text. Degrees of freedom
are labeled by discrete indices $\dirn$ and $\dirl$, respectively
transverse and longitudinal to the direction of the driving force,
$F$. Each degree of freedom interacts with its neighbors via elastic
couplings in the longitudinal direction and via viscous or similar
slip couplings in the transverse direction.}
\end{figure}

{\em Anisotropic slip model.}  Elastic models have their limits in
describing real materials.  Coppersmith argued that elastic models
with weak disorder have arbitrarily large strains, as large rare
regions with atypically low pinning result in large displacement
gradients \cite{coppersmith90&91,sncajm91}. This leads to the breakdown
of the elastic model, though the low density of such regions may
preclude its easy observation.  Simulations, imaging and noise
experiments for moderate and strong disorder also indicate the
breakdown of the elastic
model \cite{faleski96,nori96,myers99,olson_hyst01,tonomura99,troyanovski99,Maeda02}.
These calculations and observations often show a sliding state that
consists of coherent regions moving at different average velocities,
with extended defect structures along their boundaries.  Furthermore,
the dynamics is strongly anisotropic, with slip occurring
preferentially where the {shear} deformations are largest, at the
borders of channels that are on average aligned with the direction of
motion. Guided by this work, we propose an anisotropic description of
the inhomogeneous dynamics in the presence of defects.

We consider a $d=d_\parallel+d_\perp$-dimensional medium composed of
$d_\parallel$-dimensional {\it elastic channels} coupled via
interactions that allow for {\it slip} of the channels in the
remaining $d_\perp$ directions. The system is driven by a uniform
force $F$ along one of the $d_\parallel$ directions.  A cartoon of a
$1+1$-dimensional system is shown in Fig.~\ref{fig:cartoon}. We
choose to discretize space in both transverse and longitudinal directions,
using integer vectors $\dirl$ for the $d_\parallel$-dimensional
intra-layer index and $\dirn$ for the $d_\perp$-dimensional layer
index.  The local displacement along the direction of motion is
$\phi_\dirn^\dirl(t)$.  Assuming overdamped dynamics, the equation of
motion in the laboratory frame \cite{convective} is
\begin{eqnarray}
\label{model}
\dot{\phi}_\dirn^\dirl(t)
     =K\sum_{\la \dirm\ra}(\phi_\dirn^\dirm
     -\phi_\dirn^\dirl) +F
     +h_\dirn^\dirl
                 Y(\phi_\dirn^\dirl-\beta_\dirn^\dirl)
      +\sigma^\alpha_{\dirn,\dirl}\;,
\end{eqnarray}
where the dot denotes a time derivative (we have chosen to scale time
so that the damping constant is unity) and $\la \dirm\ra$
ranges over sites $\dirm$ that are nearest neighbor to $\dirl$.
The first term on the right hand side represents an elastic
intra-channel coupling of strength $K$. The third term is the
pinning force. The function $Y(x)$ has period $1$, and the $\beta_\dirn^\dirl$
are random phases chosen independently and uniformly in $[0,1)$. The
random pinning strengths $h_\dirn^\dirl$ are chosen from a probability
distribution $\rho(h)$.  Finally,
${\sigma}^\alpha_{\dirn,\dirl}$ represents an interaction (of
type $\alpha$) that allows for local slips of neighboring channels. In
this paper, we assume a linear viscoelastic ($V$)
stress-strain relation,
\begin{equation}
\label{stresstransfer}
\sigma^V_{\dirn,\dirl}=\int_{-\infty}^t dt'
    \sum_{\la {\dirp}\ra}J_{\dirp\dirn}^\alpha(t-t')
            [\phi_{\dirp}^\dirl(t')-\phi_\dirn^\dirl(t)]\;,
\end{equation}
%
%Local inertial effects are introduced 
where $\la{\dirp}\ra$ indexes the layers neighboring $\dirn$.  The
stress transfer function \cite{Fisher98,MMP00} $J_{\dirp\dirn}^\alpha$ is
generally nonlocal in space and/or time.  We focus on a simplified
version of the viscoelastic model where the local slip force is
a purely viscous coupling of strength $\eta$:
$\sigma^{\rm V}_{\dirn,\dirl}=\eta \sum_{\la 
m\ra}(\dot{\phi}_{\dirp}^\dirl-\dot{\phi}_\dirn^\dirl)$.  With this
coupling, Eq.\ (\ref{model}) is a simplified ``viscous slip''
form of the hydrodynamics
of a driven viscoelastic medium that flows in response to large-scale
shear, but responds elastically within the layers to long-wavelength
compressions \cite{boonyip80}.
Gradients of the displacement {\em
transverse} to the channels or layers correspond to {\em shear}
deformations arising from defect structures between the channels. Some
justification for modeling the slip that results from moving defects
as viscous couplings comes from work \cite{mcmks02} showing that that
the hydrodynamics of a two-dimensional crystal with free dislocations
is identical to that of a viscoelastic fluid. In contrast, gradients
of displacement {\em longitudinal} to the drive mainly yield {\it
compressional} deformations (exactly so when $d_\parallel=1$.)  The
elastic response to long wavelength compressions in fluids is
intimately related to the fact that compressions are associated with
fluctuations in the conserved density, suggesting that the 
inclusion of compressional
forces of strength $K$ is necessary to describe the driven
dynamics of systems with a conserved number of particles, such as
vortex lattices.  The coarse-grained model described by
Eq. (\ref{model}) allows us to investigate the competition between
elastic interaction and plastic flow in controlling the topology of
the nonequilibrium phase diagram.

%
%\begin{equation}
%\label{sigmaPS}
%\sigma^{\rm PS}(\{\phi_\dirn^i\})=
%     \mu\sum_{\la \dirm\ra}\sin(\phi_\dirm^i-\phi_\dirn^i)
%  \;.
%\end{equation}
%

{\em Mean-field treatment}.
%\paragraph{Viscous Model.}
One mean-field approximation for the viscous slip model is obtained by
taking the limit of infinite range interactions in both the transverse
and longitudinal directions.  Each discrete displacement then couples
to all others only through the mean velocity,
$v=N^{-1}\sum\dot{\phi}_\dirn^\dirl$, and the mean displacement,
$\overline{\phi}=N^{-1}\sum{\phi}_{\dirn}^\dirl$.  We look for
solutions moving with stationary velocity: $\overline{\phi}=vt$.
Since all displacements $\phi$ are coupled,
they can now be indexed by their disorder parameters $\beta$ and $h$, rather
than the spatial indices $\dirl$ and $\dirn$.
The mean-field dynamics
is governed by the equation
\begin{equation}
\label{MFT_viscous}
(1+\eta)\dot{\phi}(\beta,h)= K\big(vt-\phi\big)
    +F
   +\eta v+hY(\phi-\beta).
\end{equation}
When $K = 0$, the mean field velocity is determined by the
self-consistency condition $\la\dot\phi(h)\ra_{h}=v$, where the
subscript $h$ indicates an average over the distribution of pinning
strengths $\rho(h)$.  When $K\ne 0$, the mean field velocity is found
by imposing $\la\phi(h,\beta)-vt\ra_{h,\beta}=0$.

It is useful to review the case where $\eta=0$ and $K\ne 0$.  In this
limit, Eq.~(\ref{MFT_viscous}) reduces to the mean field theory of a
driven elastic medium worked out by Fisher and collaborators \cite{NF92}.
For the piecewise harmonic pinning $Y(\phi)=1/2-\phi$ for
$0\leq\phi\leq1$, no moving solution exists for
$F<F_T=\la\frac{h^2}{2(K+h)}\ra_h$.  Just above threshold the mean
velocity has a universal dependence on the driving force, with
$v\sim(F-F_T)^\beta$.  The critical exponent $\beta$ depends on the
shape of the pinning force: $\beta=1$ for the discontinuous piecewise
harmonic force and $\beta=3/2$ for generic smooth forces.  Using a
functional RG expansion in $4-\epsilon$ dimensions, Narayan and
Fisher \cite{NF92} showed that the discontinuous force captures a
crucial intrinsic discontinuity of the large scale, low-frequency
dynamics, giving the general result $\beta=1-\epsilon/6+{\cal
O}(\epsilon^2)$, in
reasonable agreement with numerical simulations in two and three
dimensions \cite{myersSIM,aamSIM}. For simplicity and to reflect the
``jerkiness'' of the motion in finite-dimensional systems at low
velocities, we use piecewise harmonic pinning.

\begin{figure}
\centering
\includegraphics[width=8.5cm]{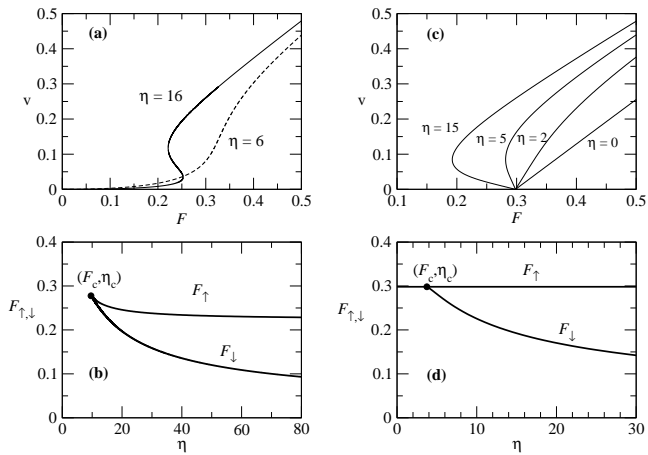}
\caption{\label{MF_viscous} Velocity curves and phase diagrams for the
mean-field slip model.  Part (a) shows the velocity $v$ vs.\ force $F$
curves for the purely viscous case for a broad distribution of
pinning, $\rho(h)=e^{-h}$, for two values of viscous coupling $\eta$.  Part (b) shows
the phase diagram from (a), with curves
curves indicating jumps in the adiabatic response (i.e.,
$dv/dF\rightarrow\infty$) when the force $F$ is increased from $F=0$
($F^\uparrow(\eta)$) and decreased from a sliding state
($F^\downarrow(\eta)$).  The critical point
$(\eta_c,F_c)\approx(9.61,0.278)$.  Part (c) shows example $v$ vs.\ $F$
curves when there is elasticity, $K=1$, for piecewise linear pinning
forces and $\rho(h)=e^{-h}$.  The curves are again continuous for
$\eta<\eta_c$, but have a depinning transition at $F_T$, with critical
behavior identical to the $\eta=0$ case.  When $\eta>\eta_c$, the
depinning is abrupt and the $v$-$F$ curves are hysteretic.  Part (d)
summarizes the $K=1$, $\rho(h)=e^{-h}$ results in a phase diagram,
indicating depinning from an initial pinned state at $F^\uparrow=F_T$
and repinning from a sliding state at a lower force $F^\downarrow(\eta)$.
The point $(\eta_c,F_T)\approx(3.74,0.298)$ is a tricritical point.  }
\end{figure}
%The phase diagram for the viscoelastic model with a broad 
%distribution of pinning forces can indeed be mapped onto that a liquid gas
%system, with $F\leftrightarrow{\rm pressure}$, $v\leftrightarrow{\rm density}$,
%and $\eta\leftrightarrow{\rm 1/temperature}$. The differential mobility
%$\mu$ 
%can be defined as $\mu=\Big(\frac{\partial F}{\partial v}\Big)_{\eta}$.
%The low-mobility phase 
%corresponds to the liquid and the high-mobility phase to the gas. 
%The mobility diverges at the critical point.}

% 

When $\eta > 0$, the nature of the depinning differs qualitatively
from the $\eta=0$ case in that hysteresis in the dynamics can take
place.  This can be shown starting from the $\eta=0$ solutions
$v_{\eta=0}(F)$, with $v_{\eta=0}$ a well-defined function of $F$.
The solution for general $\eta$ can be found by substituting the
effective drive force $G=F+\eta v$ for $F$ in the $v_{\eta=0}(F)$
relation and scaling velocity down by $1+\eta$.  The linear
transformation $F = G-\eta v$ then gives the general $v(F)$ curve. One
result is that when $1+\eta^{-1}<\max_F (dv_{\eta=0}/dF)$, the $v$ vs.\
$F$ curve is multivalued.  Two cases can now be considered. The
viscous-elastic case $K\ne 0$ has quite different behavior from the
purely viscous case $K = 0$, due to the distinct self-consistency
conditions imposed by elasticity (or conservation in the flow
direction.)

For the purely  viscous case $K=0$ (the $\tau=0$ limit of the model in
\refref{MMP00}), periodic solutions of period $T(h,G)$ are found for a
single particle moving in a pinning force of strength $h$ under the
effective drive $G$.  Self-consistency implies that
$v=\la [T(h,G)]^{-1}\ra_h$, giving
\begin{equation}
\label{F_viscous}
%F=-\eta v+\Big\la\frac{h}{2}~\frac{1+e^{-h/(1+\eta)v}}
%       {1-e^{-h/(1+\eta)v}}\Big\ra_h\;.
v = \frac{1}{1+\eta}\int_0^{2G} dh\,\rho(h)
h\left[\ln \left(\frac{G +h/2}{G-h/2}\right)\right]^{-1}\;.
\end{equation}
The behavior now depends on the shape of the disorder distribution
$\rho(h)$. For narrow distributions, that vanish below a value $h_{\rm
min}>0$, no moving solutions exist for $F<h_{\rm min}/2$.  In the
extreme case of uniform pinning, $\rho(h)=\delta(h-h_0)$, the
depinning is hysteretic for all nonzero values of $\eta$, with a jump
from $v=0$ to finite $v$ at $F=h_0/2$.  For the more general broad
distribution with nonvanishing support at $h=0$, no pinned solution
exists for $F>0$: while the bulk of the degrees of freedom are pinned
at small $F$, weakly pinned $\phi$ can respond to small drives.
Typical $v$ vs.\ $F$ curves for $\rho(h)=e^{-h}$ are shown in
Fig.\ 2(a).  The lines in Fig.\ 2(b) represent the critical forces
where there is a macroscopic jump in the average velocity, indicating
dynamic hysteresis.  For $\eta > \eta_c$ and
$F_\uparrow>F>F_\downarrow$ ``fast'' and ``slow'' sliding states can
coexist.  The critical point $(\eta_c,F_c)$ in Fig.~2(b) is in the
universality class of the liquid-gas transition (and of the
field-driven random field Ising model \cite{MCMKD02}.) Near the
critical point the mean velocity has universal scaling,
$v-v_c\sim(\eta-\eta_c)^{1/\delta}\sim(F-F_c)^\beta$, with
$v_c=v(F_c,\eta_c)$, and $\beta_{MF}=1/2$, $\delta_{MF}=3$.  The point
$(\eta_c,F_c)$ survives in
finite dimensions \cite{MCMKD02}, with exponents distinct from their mean
field values.

For finite long-time elasticity, i.e., when $K\not=0$, the behavior
changes dramatically. The elastic forces or particle conservation
enforce a uniform time-averaged velocity for all degrees of freedom.
In this case the long-time uniform-$v$ solution to
\eqref{MFT_viscous} is $\phi(h,\beta,t)=vt+\tilde{\phi}(h,\beta,t)$
with $\tilde{\phi}(h,\beta,t)$ of period $1/v$ in $t$.
The self-consistency
condition is $\la\tilde{\phi}(\beta,t)\ra_{\beta,h}=0$.
The explicit solution can be obtained for the piecewise
linear pinning force.  (In the mean-field limit the nature of the
phase diagram depends some on the shape of the pinning potential.)
There is now a pinned phase that is stable \cite{stabilitynote} for
$F<F_T=\la\frac{h^2}{2(K+h)}\ra_h$.
One then solves
\begin{equation}
\label{F_viscous+elastic}
\delta F=\frac{\eta_c-\eta}{1+\eta_c}v
  +\Big\la\frac{h^2}{(K+h)}\frac{1}{e^{(K+h)/[(1+\eta)v]}-1}
  \Big\ra_h\;,
\end{equation}
to find the mean velocity in the sliding
state ($0<\delta F\equiv F-F_T$),
where $(\eta_c+1)^{-1}=\la\frac{h^2}{(K+h)^2}\ra_h$.  The
velocity-force curves and a phase diagram are shown in
Fig.\ 2(c,d) for $\rho(h)=e^{-h}$.
There is now a {\it tricritical
point} at $(\eta_c,F_c=F_T)$.
For $\eta<\eta_c$, a continuous depinning transition at $F_T$
separates a pinned state from a sliding state with {\em unique} velocity
$v\sim(1+\eta_c)\frac{F-F_T}{\eta_c-\eta}$, giving $\beta=1$ in
MFT.  In finite dimensions, this transition is likely to remain in the same
universality class as the depinning of an elastic medium ($\eta=0$):
numerical studies and analysis by Schwarz and Fisher \cite{SF02} of a model
with local viscous-like terms
show that there is no hysteresis for small slip coupling
and that the depinning transition exponents are the same as
without slip coupling.  In our mean-field example, the
linear response diverges at $\eta_c$,
$v(\eta=\eta_c)\sim 1/\ln(F-F_T)$.  For $\eta>\eta_c$
there is hysteresis with coexistence of stuck and sliding
states.

{\em Relations to other models}. In the absence of viscous coupling
($\eta=0$), \eqref{model} reduces to the conventional
$d_\parallel$-dimensional phase-only model of driven CDWs.  When $K=0$
and $\eta$ is finite, the model describes shear deformations in a
viscous fluid in a disordered background \cite{MMP00}.

The mean-field limit of the anisotropic slip model for finite $K$ maps
onto a periodic-pinning version of
the infinite-range version of the stress overshoot model studied
by Schwarz and Fisher \cite{SF01,SF02}. The dynamics of crack fronts in
brittle materials can be dominated by local inertial effects, in which
the motion of a crack segment creates a transient stress on other
segments, the stress overshoot, resulting in a net forward motion
beyond the local minimum of the pinning.  In the infinite-range limit
the stress transfer yields a global coupling of strength $M$ between
the mean displacements at two different times, analogue to the
effective force per unit velocity $\eta$ in our model.  The behavior
we conjecture for finite dimensions is identical to that obtained by
Schwarz and Fisher.

The viscous model ($\eta\not=0$) with finite $K$ is also closely
related to a model of sliding CDWs
that incorporates the coupling of the CDW to normal carriers by adding
a global velocity coupling \cite{littlewood88,levy92,levy94}
to the conventional Fukuyama-Lee-Rice
model. The precise relationship is obtained
by considering a mean-field limit $d_\perp\rightarrow\infty$ with
fixed $d_\parallel$. In this limit, each degree of freedom is
coupled equally
by slip to all other layers, but elastically
coupled only to its nearest neighbors in the longitudinal direction.
The equation of motion is then
\begin{equation}
\label{MFT_levy}
(1+\eta){\dot\phi}^i_\dirn=
   K\sum_{\la j\ra}(\phi_\dirn^j-\phi_\dirn^i) 
        +F+\eta v\nonumber
   +h_\dirn^iY(\phi_\dirn^i-\beta_\dirn^i)\;.
\end{equation}
\eqref{MFT_levy} is identical to the equation of motion of a purely
elastic CDW with friction $1+\eta$ and an effective driving force
$F+\eta v$.  One can then obtain the velocity-force characteristics
consistent with \refref{levy94} simply by translating the analytic
results \cite{NF92} for CDWs. Near threshold, we obtain $v\sim(F+\eta
v-F_T)^\beta$, with $\beta=1-(4-d_\parallel)/6+{\cal
O}[(4-d_\parallel)^2]$.  The transition is always hysteretic: the
medium depins at $F^\uparrow=F_T$ when the force is ramped up and
repins at a lower value, $F^\downarrow\approx
F_T-\Big(\frac{1}{\beta}-1\Big)
\Big(\frac{\beta\eta}{1+\eta}\Big)^{1/(1-\beta)}$, when the force is
ramped down from the sliding state. This hysteresis for all $\eta > 0$
appears to be a consequence of the global nature of the coupling, in
contrast with the results of Schwarz and Fisher for local slip
coupling \cite{SF02}.

One can include phase-slip couplings $\sigma^{\rm PS}$
periodic in $\phi$. The anisotropic model can describe the driven
dynamics of intrinsically layered systems, such as vortex lattices in
cuprate materials. Hysteresis depends on the strength of the
correlations in the (weak-coupling) layered direction, as seen in
simulations \cite{olson_hyst01}. An anisotropic CDW model
was studied in this
context by Nattermann and Vinokur \cite{NV97}. These authors modeled
the transverse slip as a nonlinear coupling that is periodic in the
local CDW phase differences; this allows for phase slips, i.e., the
collapse of the CDW amplitude between coherent
regions \cite{strogatz88&89}.

{\em Summary.} We have proposed that the plastic flow of extended
systems in a disordered background can be modeled using a general
anisotropic approach that includes many specific physical systems as
special cases. We have discussed the behavior of this model in a
mean-field limit.
When the degrees of freedom are allowed to move at distinct velocities,
a pinned phase is generally not present, though there is hysteresis in
the current-drive relation.
Even when fluid-like shear takes place, particle conservation
gives a sharp depinning transition in flow that takes place along
channels.
One of our future goals is to establish the connection between the 
parameters of the model and physical parameters of a given experimental system.
This work has been supported in part by the National Science Foundation via
grants DMR-9730678 and DMR-0109164.


\begin{thebibliography}{99}
\bibitem{footcontinuous} 
Periodic 1D models
under a.c. drive can have discontinuous non-hysteretic depinning curves
(See \refref{aamSIM}.)
\bibitem{convective}
We drop the convective derivative in the equation of motion 
as it does not contribute in the mean field limit.
\bibitem{stabilitynote} This stability can be history dependent
or vanish at finite driving rates, as discussed
in Ref.\ \onlinecite{SF01}.
\end{thebibliography}
\end{document}